\documentclass{eptcs}
\providecommand{\event}{TFPIE 2019}
\usepackage[noxcolor]{pstricks}
\usepackage{mweights}
\usepackage{fontaxes}
\usepackage{pst-node}
\usepackage{graphicx}
\usepackage{caption}
\usepackage{subcaption}
\usepackage{stfloats}
\usepackage{multido}
\usepackage{setspace}
\usepackage{textcomp}
\usepackage{amssymb}
\usepackage{amsmath}
\usepackage{alltt}
\usepackage{tikz}
\usepackage{pgfplots}
\RequirePackage{doi}
\usepackage{hyperref}

\newcommand{\cw}{\texttt{CodeWorld}}
\newcommand{\hdp}{\texttt{HtDP}}

\newcommand{\racket}{\texttt{Racket}}
\newcommand{\comment}[1]{}

\title{Using Video Game Development to Motivate Program Design and Algebra Among Inner-City High School Students}

\author{Marco T. Moraz\'an
\institute{Seton Hall University}
\email{morazanm@shu.edu}}
\def\titlerunning{Motivate Program Design Among Inner-City High School Students...}
\def\authorrunning{M. T. Moraz\'an}

\begin{document}
\maketitle

\begin{abstract}
Introducing inner-city high school students to program design presents unique challenges. The typical assumptions of an introductory programming course, like students understand what variables and functions are, may not be safe. Therefore, asking students to define functions as part of the program design process may be an overwhelming task. Many students do not understand that a function is an abstraction over similar expressions and that parameters represent the differences among these expressions. This articles presents a novel approach to teaching program design to high school students while simultaneously reinforcing high school algebra. The approach is based on a design recipe to help students develop the abstractions that lead to functions. Using a bottom-up approach, students are taught how to abstract over similar expressions. They are then taught how to use high school algebra concepts, like compound functions and function composition, to also design functions. In addition, the article also presents empirical data collected from students to measure their reaction to the course. For the students in the course, the empirical data suggests that high school algebra concepts are successfully reinforced and that students feel they become better problem solvers, find programming intellectually stimulating, and walk away with an interest in programming.
\end{abstract}

\section{Introduction}
In a programming course for beginners, picking a starting point may seem obvious. Many textbooks start by explaining some primitive types, some basic syntax, and some control structures (e.g., \cite{Eck,Sedgewick}). Others, more appropriately, start with expressions (e.g., \cite{SICP,HtDP,HtDP2}). Invariably, however, most textbooks then move to defining functions/methods. The assumption is that functions come naturally to students especially if they have taken a course in high school algebra (or higher). It would seem to be a natural and safe assumption that such students know what variables/parameters and functions represent. After all, high school algebra students are drilled with exercises that required them to plug-in values to a function. That is, students learn to substitute the references to the independent variable with a value. Consider the following typical quadratic equation found in any high school algebra book:
\begin{alltt}
     y = \(x\sp{2}\) + 3x - 10
\end{alltt}
Students are commonly asked, for example, to find the value of \texttt{y} when \texttt{x} is 10. Most students then proceed to substitute \texttt{x} with 10 on the righthand side of the equation to obtain \texttt{$10^2 + 3*10 - 10$}. This correctly leads students to the conclusion that the value of \texttt{y} is 120. Does this mean that students understand functions?

It turns out that many beginning students do not understand what a function is. Put differently, they do not understand what a function represents. To functional programmers and to mathematicians, a function is an abstraction that captures a common computation pattern. To a programming beginner, it is no such thing. Instead, a function is something (no clear definition) in which you substitute a variable for a value. Therefore, it is not shocking that beginning students stumble when asked to write their own functions. We are asking them to develop something that has always been provided to them. More importantly, we are asking them to define something that they do not comprehend. The bottom line is that beginning students usually do not understand abstraction over expressions when they first walk into the classroom.

Given a room of eager beginning programming students that learned high school algebra by rote, how do you explain to them how to write a function? This is a dilemma the author faced with a group of 11$^{\texttt{th}}$ graders taking their first-ever programming course in the summer of 2019. This group of students did not understand abstraction at any level, having difficulty to express that a variable represents a value, never mind that it represents a difference between similar expressions. In fact, students insisted in always using and thinking in terms of concrete values instead of in terms of variables. The solution the author converged on is to first teach students about expressions using concrete values and then teach them how to abstract over similar expressions to create functions. Instead of using a top-down approach to teach them how to design solutions to problems (i.e., functions), a bottom-up approach is adopted. Auxiliary functions are developed first because it is usually easier to write sample expressions using concrete values--something the students crave. Once auxiliary functions have been developed, students feel more comfortable developing the functions that make use of them. The results were nothing short of encouraging. By the end of the course, not only did students understand what a function is, but also built on that knowledge to create a functional video game.

The article is organized as follows. Section \ref{RW} presents related work. Section \ref{back} briefly discusses the student background. Section \ref{CO} presents a brief overview of the course. Section \ref{Exp} outlines how expressions are discussed and abstracted over in class. Section \ref{Ex} presents the video game developed with students building on the new knowledge they have acquired. Section \ref{Eval} presents student feedback. Finally, Section \ref{Concl} presents concluding remarks and directions for future work.

\section{Related Work}
\label{RW}
It is not uncommon for a high school student's only exposure to functions to have been in a Mathematics course. Typically in the United States, the common denominator for all students is high school algebra. Some high school algebra textbooks (e.g., \cite{Schaum}) start with so-called four fundamental operations: addition, subtraction, multiplication, and division. These operations are never identified as functions. From here, algebraic expressions are introduced as a combination of ordinary numbers and letters. At the beginning, these ``letters" are not called variables. A great deal of work is done with expressions before introducing functions. Functions are introduced as relations that associate an element of the domain with exactly one element of the range, but are never associated with abstracting over expressions. Similarly, the work presented in this article builds on first developing expressions and then introducing functions. In contrast, however, functions are explicitly introduced as an abstraction over expressions and as entities that compute values. The writing of unit tests is used to drive home the idea that functions compute values. More specifically, functions compute the same values that are obtained from sample expressions.

Efforts to get high school students engaged in programming introduce them to the development of functions. \texttt{Bootstrap} \cite{BS,BS2}, for example, integrates introductory programming and high school algebra. In the \texttt{Bootstrap} curriculum, function definitions are introduced after observing that a program may have expressions that are not identical, but are very similar. Functions are described as taking inputs and producing output. Students are then told that programmers may want functions that are not provided by the programming language to abbreviate expressions that are very similar. Their examples tell students what the input and output are and then proceed to write examples to identify and give the observed differences (i.e., the inputs to the function being developed) variable names. At this point students can define a function by copying the similarities and using the variable names in place of the differences. Similarly, the work described in this article closely couples programming and high school algebra. The approach described first has students identify differences between similar expressions and use variables to abstract them away. In contrast, however, the work presented in this article does not prescribe the inputs and output of a function beforehand. Instead, students are asked to first write sample expressions that demonstrate how to compute a value. These expressions are used by students to discover the input variables (i.e., parameters) a function needs--one for each difference identified. Students use these to write a signature, purpose statement, and function header. Before defining the body of a function, students write unit tests using their sample expressions. After this step, students proceed to define the function's body keeping the similarities in their sample expressions and using the function's parameters in place of the terms that vary. The difference between \texttt{Bootstrap} and the approach described in this article may appear subtle to some readers, but in practice it allows students to be active participants in discovering the inputs to a function. In this manner, students are able to understand that a function is an abstraction that captures a computation pattern born out of similar expressions that use concrete values.

The textbook, \textit{How to Design Programs} (\hdp) \cite{HtDP}, also integrates high school algebra with programming. Functions are introduced as the result of completing values in a table for, say \texttt{x}, the independent variable and, say \texttt{y}, the dependent variable. Once a table is populated with values, students are asked to find an ``expression" that determines any element, \texttt{y}, for a given value of \texttt{x}. For example, for a given table, students may determine that \texttt{$y = x^2 + 10x + 5$}.  \hdp \ correctly claims that this notation is misleading, given that it does not properly emphasize the fact that this is a function definition that may be used to compute any desired \texttt{y} value in the table. It then uses this observation to motivate the need for students to learn more syntax to define functions. The assumption is that students arrive in the classroom with the necessary skills to extract functions from sample tables. This assumption is not always safe, and students need to be guided through a series of steps that allows them to create the abstraction (i.e., the function). The work presented in this article builds on \hdp \ to help students create the abstraction. In essence, it expands the first step, \emph{problem analysis}, of any design recipe found in \hdp. Students write sample expressions using concrete values to compute a desired result as part of the first step of development. This is tantamount to creating a table, but a key difference is that it forces students to first think about how a value is computed instead of simply filling in values in a table. Students use the expressions they write to discover the inputs needed for a function and the expression needed in the function's body. That is, they develop the abstraction from sample expressions, not sample values. Furthermore, the values of the sample expressions are used to simplify the development of unit tests.

Another effort that integrates high school algebra with programming for beginners is \cw~\cite{cwguide}. \cw \ presents a programming environment, built on \texttt{Haskell}, in which you can create drawings, animations, and video games. Students are taught using high school algebra concepts such as expressions, variables, functions, domain, range, and function composition. In addition, students are introduced to types to describe expressions and functions. An expression is described as code to describe a value. A function is described as a relationship that associates input values with a specific result. Variables are introduced as a mechanism of abstraction motivated by the \textit{Don't Repeat Yourself} principle. Instead of repeating the same expression within an enclosing expression, students are encouraged to define a variable to represent the expression. In contrast to variables that represent a single value, functions are introduced as capturing a general idea that arises from differences in similar expressions. If expressions are similar, the \cw \ methodology suggests naming variables for the differences, using these variables to write a new function, and rewriting the original expressions using the new function. Similarly, the work described in this article starts students with expressions and motivates functions as abstractions over expressions. In contrast, the work described in this article presents students with a complete set of steps to design functions\footnote{The design sections of the \cw \ guide remain to be written.}. For example, unlike the \cw \ methodology, students are encouraged to define variables for the value of expressions, to write a signature (i.e., domain and range) and a purpose statement for a new function, and to write unit tests using the defined variables for the value of sample expressions and the new function.

Efforts under the mantle of computational thinking have also tackled the problem of teaching beginners. Michaelson, for example, argues that problem solving should be driven by abstraction from concrete instances of a specific problem \cite{Mich1,Mich2}. By identifying the underlying regularities in differences among concrete calculations, students can determine the types and variables needed to develop a generalized expression. These efforts are decoupled from the syntax of any programming language. Likewise, the work described in this article has students identify variables and their types from expressions using concrete values. In contrast, the expressions are written using the syntax of a programming language. The goal is to facilitate the transition from problem analysis to coding. Furthermore, the concrete expressions developed are used in unit tests which facilitates communicating to others (e.g., fellow students or an instructor) how a value is computed--a fundamental goal of programming \cite{PPL}.

\section{Student Background}
\label{back}
All students are participants in the \emph{Upward Bound Program} (\texttt{UBP}) \cite{UBP} implemented at Seton Hall University. The \texttt{UBP} is funded under the auspices of the Federal TRIO Program (\texttt{TRIO}) \cite{TRIO}. \texttt{TRIO} is comprised of outreach and student services programs for individuals from disadvantaged backgrounds (e.g., low-income individuals, first-generation college students, and individuals with disabilities). The goal is to help these individuals to progress academically from middle school to post-baccalaureate programs.

\texttt{UBP} provides support to individuals in their preparation for college entrance in an effort to increase the rate at which participants complete secondary education and enroll in and graduate from institutions of postsecondary education. All \texttt{UBP}s must provide instruction in math, laboratory science, composition, literature, and a foreign language. Programming is offered in addition to these topics at Seton Hall University. Students must have completed the 8th grade, be 13--19 years old, and have a need for academic support in order to pursue a program of postsecondary education. The program requires that two-thirds of the students in a project be both low-income and potential first-generation students. The remaining one-third may also include students who have a high risk for academic failure.

All 15 students in Seton Hall's 2019 programming course are 11$^{\texttt{th}}$ graders from inner-city high schools in New Jersey (e.g., from East Orange, Newark, and Irvington). The average age of the students is 16.13 ($\tilde{x}$=16 and $Mo$=16) and the age range is [15..18]. The majority of the students, 67\%, are females and 33\% are males. Ethnically, the students are 87\% African Americans (with 1 student identifying as Haitian and 1 student identifying as Jamaican). Virtually all students, 93\%, stated having a desktop or a laptop computer at home. The majority, 60\%, stated having no programming experience at the beginning of the course. The 40\% identifying themselves as having programming experience stated that this experience is in \texttt{Java} or \texttt{Scratch}. All of the students have successfully completed a high school Algebra II (or higher) Mathematics course.

\section{Course Overview}
\label{CO}
The initial goal was to deliver a course that uses video game development to motivate students to learn the basics of programming by design \cite{HtDP,HtDP2}. The course aimed to be a subset of a similar course developed for first-year undergraduates \cite{mtm22} that includes:
\begin{itemize}
  \item Primitive types (i.e., numbers, strings, Booleans, and images)
  \item Definitions
  \item Conditionals
  \item Compound data of finite size
  \item The Design Recipe
  \item Compound data of arbitrary size (i.e., lists and natural numbers)
  \item Functional abstraction
\end{itemize}
The course started well, and students were excited about creating and manipulating images (similar to the approach taken by others \cite{Bloch,cwguide}). Although defining constants did not present a challenge, the course quickly ran into trouble when students were asked to define functions.

Students felt unable to write functions on their own. They expressed not understanding what they are being asked to do. This was rather surprising because, as indicated above, 100\% of the students have already successfully taken an Algebra II course. That is, students have been exposed to the following concepts: variables, functions, domain, range, plugging-in values, problem-solving, the Cartesian coordinate system, graphing, compound functions, and function composition. Despite this exposure, students had the following questions and comments:
\begin{itemize}
  \item \emph{Where do functions come from?}
  \item \emph{Why are you not giving us the functions?}
  \item \emph{I can't write something I don't understand.}
  \item \emph{What does it mean that a function computes a value?}
  \item \emph{What do you mean by the input to the function?}
\end{itemize}
These stumbling blocks arose even with problems that most instructors would consider fairly straight-forward, such as finding the area of a rectangle.

Given these challenges, it was clear that the course needed to be redesigned. The course now directly reinforces concepts studied in high school Algebra and shows the students how Algebra is used in problem solving and computer programming. In addition, the course still aims to keep students enthusiastic and engaged by developing a video game. The topics covered by the course now are:
\begin{itemize}
  \item Primitive types (i.e., numbers, strings, Booleans, and images)
  \item Expressions
  \item Abstraction over expressions, function definitions, domain, and range
  \item Compound functions and conditionals
  \item The Design Recipe
  \item Compound data of finite size
\end{itemize}

\begin{figure}[t]
 \begin{center}
  \includegraphics[scale=0.3]{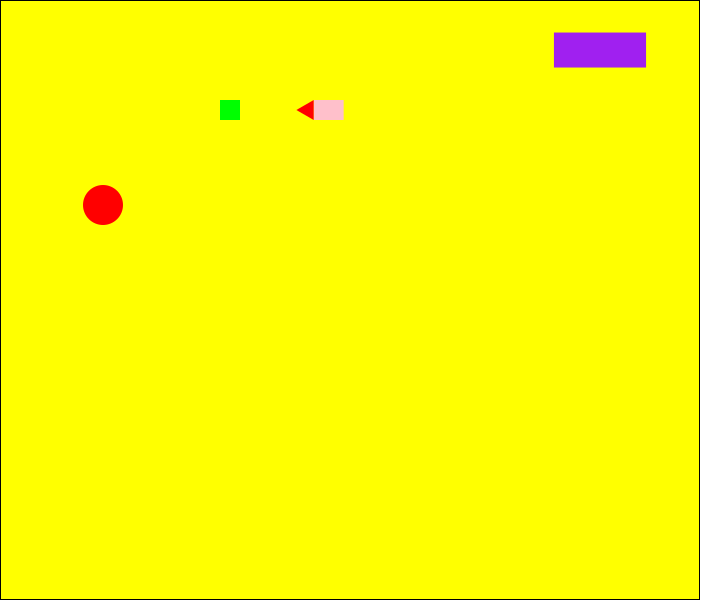}
 \end{center}
 \caption{The Rocket Game} \label{game}
\end{figure}

A rendition of the video game developed in class is presented in Figure \ref{game}. It consists of a rocket that may consume good (green square) or bad (red circle) fuel and a fuel level (purple box). The player can change the direction the rocket travels in by using the arrow keys. Every time the clock ticks, the rocket moves in its current direction by a constant amount and the fuel level decreases by a constant amount. If the rocket consumes good fuel, the fuel level goes up by some constant amount without surpassing a maximum limit (say, 10), and the good fuel moves to a random location within the scene. If the rocket consumes bad fuel, the fuel level goes down by some constant amount without decreasing beyond empty (say, 0), and the bad fuel moves to a random location within the scene. The game ends when the fuel level is empty.

The programming language used is Racket's Beginning Student Language (\texttt{BSL}) \cite{HtDP,HtDP2}. \texttt{BSL} was chosen as the language of instruction because it is pure (i.e., does not support assignment) and has error messages tailor-made for beginners. The video game is developed using the \texttt{universe} library \cite{univ}. Development proceeds using a bottom-up approach. That is, the simplest and, perhaps, most obviously needed functions are developed first. These functions are then used to create more complex functions. For example, first, functions to move the rocket in every direction are developed, and then these functions are used to develop a function to move the rocket when a key is pressed. This bottom-up approach was chosen to simplify the introduction of function definitions and to satisfy the student's need to use concrete values.

\section{The Nature of Functions}
\label{Exp}
To answer student concerns about where functions come from, they are first carefully introduced to expressions. Students, in essence, are asked to write examples of how to compute different instances of the same abstract value (e.g., an area or a circumference). One important goal is to have students grow tired of repetitive typing. That is, students are given the opportunity to come to the realization that there must be a better way to write expressions that, in essence, compute the same (abstract) value. This better way, of course, is to write a function.

To achieve this, an abstraction step is needed. Students are taught to identify the differences between two or more similar expressions. These differences are what may vary from one expression to the next, and students are asked what is something that varies called. After some class discussion, students converge on the term \emph{variable}. Students are taught that functions, therefore, need variables to represent the differences between similar expressions. These variables are the input to the function and are called parameters.

With this realization in mind students are given the following design recipe to develop a function:
\begin{enumerate}
  \item Define variables to store the value of sample expressions.
  \item Identify the differences among the sample expressions.
  \item Give each difference a parameter name.
  \item Identify the type of each parameter and the return type for the function's signature.
  \item Identify the purpose of the function.
  \item Write a function header.
  \item Write tests illustrating how the function ought to work.
  \item Write the body of the function.
  \item Run the tests and redesign if necessary.
\end{enumerate}
Observe that each step has a specific outcome. This means that both students and instructors can verify whether or not a step has been properly completed. Thus, instructors can provide a student with concrete feedback when a step is not properly completed. Furthermore, students are instructed that a primary goal of programming is to explain to others how a problem is solved \cite{PPL}. To foster this goal, they ought to correctly complete each step of this design recipe.

The first 3 steps outline for students part of what is called ``problem analysis" in \hdp. Step 1 has students define variables to store the values of sample expressions to be used for testing. These expressions must demonstrate how a value is computed. This requires that students plan how to compute a value. It is important to guide students to write expressions in a consistent manner. For example, for the area of a rectangle all expressions must be written as \texttt{width x height} or as \texttt{height x width}, but not both. As a result of this step, students realize that the different expressions they develop have a lot in common and, therefore, the similarities ought to be captured in a function. Step 2 has students explicitly identify the subexpressions that are different from one expression to the next. These subexpressions need to be represented by variables that will be parameters. Step 3 has students develop a parameter name for each difference. Students are strongly encouraged to select descriptive names and avoid simplistic names such as \texttt{k} or \texttt{x}.

Steps 4 and 5 have students write the signature and purpose statement of the function. This introduces students to type-oriented programming. At this level, two things are very important. The first is that students not face cryptic type-error messages. Cryptic error messages tend to discourage beginners and this is why \racket's student languages and others have a tailored-made error-messaging system for beginners \cite{Crestani,HtDP,Hsia,Munson,Schiliep}. The second is for students to become familiar with writing down types without being inhibited from quickly building prototypes even if their code is ill-typed. In the context of this course, the types are needed to properly design the function and are not seen as a hurdle that must be overcome to get programs to run. Furthermore, the signature of a function is associated with the concepts of the domain and range of the function that students have studied in high school algebra.

Step 6 has students write the header of the function. This requires naming the function and using the parameter names developed in Step 3. After writing the header of the function, Step 7 has students write examples of how the function ought to work. These are written in the form of unit tests using Racket's test engine (specifically, \texttt{check-expect}, \texttt{check-random}, and \texttt{check-within}) \cite{RTEST}.  If there is variety in the data, for example, students must include at least one test for each variety. The tests must use the function being developed, and there must be a test using each of the variables defined in Step 1. It is emphasized to students that function arguments must be provided in the order expected by the function header. Students, of course, may have to write more than one test per variety to achieve a thorough coverage of possible inputs. It is important to highlight to students that the use of the function is much shorter (and more elegant) than using the expressions written to define the variables in Step 1. In addition, students must also write tests using concrete values that differ from those found in the sample expression used in Step 1. This is necessary to drive home the point that the function works for more than just the values used in the sample expressions.

For Step 8, students write the expression for the body of the function. This expression must structurally look the same as the sample expressions written for Step 1. The differences identified in Step 2, however, are replaced with the corresponding parameter names developed for Step 3. It is highlighted to students how substituting each concrete value provided to the function in the tests results in the expressions developed in Step 1.

Step 9 has students run their tests. Students must make sure that their tests are thorough and have every bit of code covered. If there is code not tested, students must add tests. In \texttt{DrRacket}, this step is simplified given that untested code is automatically highlighted.

To illustrate how students are first introduced to the development of functions, consider starting with a simple computation that ought to be familiar to all students that have taken high school algebra. For example, students are asked to consider how to compute the area of a rectangle. Most students are able to state that the area of a rectangle is given by the product of its length and its width.

\begin{figure}[t]
\begin{alltt}
; Sample Expressions
(define AREA1 (* 10 5))  ; for length of 10 and width of 5
(define AREA2 (* 50 2))  ; for length of 50 and width of 2
(define AREA3 (* 4 25))  ; for length of 4 and width of 25

; Tests
(check-expect (rect-area 10 5) AREA1)
(check-expect (rect-area 50 2) AREA2)
(check-expect (rect-area 4 25) AREA3)
(check-expect (rect-area 2 7) 14)
(check-expect (rect-area 50 5) 250)

; \(\mathbb{R}\sb{\geq0}\) \(\mathbb{R}\sb{\geq0}\) \(\rightarrow\) \(\mathbb{R}\sb{\geq0}\)
; Purpose: To compute the are of a rectangle from the given length and width
(define (rect-area length width)
  (* length width))
\end{alltt}
\caption{The Function to Compute the Area of a Rectangle.} \label{area}
\end{figure}

For Step 1 of the design recipe, students develop several variable definitions representing different areas. A representative sample is displayed in Figure \ref{area}. Observe that students write all the sample expressions in the same manner: \texttt{(* length-number width-number)}. In this manner, students easily observe that the three expressions share many common elements: \texttt{(}, \texttt{*}, and \texttt{)}. Therefore, these expressions suggest the development of a function to capture the similarities in a single location.

In step 2, the students identify that the values for length (i.e., 10, 50, and 4) and for width (i.e., 5, 2, 25) may vary from one expression to the next. Having identified the differences, students may name the parameters \texttt{length} and \texttt{width} in Step 3.

To develop step 4, students are asked to identify the data type of each parameter. Commonly, students will first answer a number. They are then asked if the width can be -32.1 or if the length can be -9. Most students are quick to answer no, and it is pointed out to them that it does not suffice to simply state that \texttt{length} and \texttt{width} are numbers. After some class discussion, students arrive at the conclusion that these variables must represent non-negative real numbers, denoted by $\mathbb{R}_{\geq 0}$. Attention is now turned to the range of the function or, equivalently, the type of the returned value. It does not take long for students to realize that the returned type is also $\mathbb{R}_{\geq 0}$. Step 5 is fairly straightforward for students in this case, and they state that an area of a rectangle is computed from its length and width. These observations lead to the signature and purpose statement displayed in Figure \ref{area}.

For step 6, students must write the function header using a descriptive name. It is common for students to suggest naming the function \texttt{area}. Although the suggested name is descriptive, it may not be the most adequate. Students are asked if the function can be used to compute the area of a circle. This leads them to realize that it is a good idea to make the name more descriptive by somehow specifying the geometric shape. To reduce the amount of typing, a compromise is reached to use \texttt{rect-area} instead of \texttt{rectangle-area}. This is not an unreasonable compromise, and high school students are pleased to have to type less (even though we are just speaking of a few characters).

The tests written for step 7 require students to use the proposed function giving it as input the concrete values used in the expressions developed for Step 1 and to use the defined variables from Step 1 as the expected result. In Figure \ref{area}, the first three tests are of this type. In addition, students write tests using concrete values that differ from those used in Step 1 to demonstrate that the function works for arbitrary values. In Figure \ref{area}, the last two tests are of this type.

The body of the function is written in Step 8 by using the expressions for the variables of Step 1. Instead of using specific values for the differences (like 10, 50, and 4 for the length and 5, 2, 25 for the width), the variables \texttt{length} and \texttt{width} from Step 3 are used. Finally, to satisfy Step 9, students run their tests. If students have no syntax errors, all code is tested, and none of the tests fail, then students can be relatively confident that they successfully designed the function to compute the area of a rectangle. Otherwise, they must correct their design and rerun. To achieve this, students are advised to review each of their answers for the steps of the define recipe. For example, if a student wrote this test:
\begin{alltt}
     (check-expect (rect-area 10 100) AREA1).
\end{alltt}
The test fails and the student must either revise their answer to Step 1 by changing the value used for the width to 100 or revise their answer to Step 7 and revise the above test to:
\begin{alltt}
     (check-expect (rect-area 10 5) AREA1).
\end{alltt}

\section{Video Game Development}
\label{Ex}
A new design recipe that teaches high school students that a function is an abstraction over similar expressions does not suffice. These students need further motivation. The domain chosen to get students personally engaged is the development of a video game. The instructor's real goal, of course, is not the development of a video game. Instead, the real goal is to develop problem solving and critical thinking skills along with the reinforcement of lessons from a high school Algebra course.

This section outlines how the design recipe from the previous section is used to reinforce high school Algebra lessons in a problem solving scenario. Specifically, it outlines how students are taught the relevance of relational functions, compound functions, and function composition. To get students started with a video game they are introduced to structures to glue together values that are related. For example, a coordinate in the (two dimensional) Cartesian plane has an \texttt{x} and a \texttt{y} value. These two values are glued together in a structure called \texttt{posn}. For instance, \texttt{(make-posn 1 2)} is a coordinate with \texttt{x = 1} and \texttt{y = 2}. Students  are now encouraged to think about structures as a single value. That is, students are encouraged to think of a defined structure as a new type of data.

The presented examples assume the following data definition for the video game's world:
\begin{alltt}
A world is a structure, (make-world rocket dir flevel gfuel bfuel), where
       rocket is a posn
       dir is either "right," "left,", "up," or "down"
       flevel \(\in \mathbb{R}\sb{\geq0}\)
       gfuel is a posn
       bfuel is a posn
\end{alltt}
The rocket, the good fuel, \texttt{gfuel}, and the bad fuel, \texttt{bfuel}, are coordinate structures inside the video game's scene of dimensions \texttt{WIDTH} and \texttt{HEIGHT}. The \texttt{flevel} represents the rocket's fuel level. Finally, \texttt{dir} represents the direction the rocket is moving in.

For a seasoned programmer or a student with some programming experience, the exercises presented in this section may seem simplistic. Clearly, students well-versed in high school algebra can solve more complex problems. It is important, however, to keep in mind the audience this course targets. As outlined in Section \ref{back}, the target audience is inner-city high school students who have a high risk for academic failure and mostly have no programming experience. In addition, it is also important to recall the difficulties, as described in Section \ref{CO}, that students expressed. Overall, the students are not well-versed in high school algebra and need to be motivated to learn high school algebra and programming. Hence, the course needs to start with very basic examples. This is not to suggest that these students lack talent or are incapable of developing enthusiasm for academic endeavors. On the contrary, through the presented exercises many of these students were able for the first time to see high school algebra in practice, develop enthusiasm for the material, and learn about programming.

\subsection{Relational Functions}
\label{relf}

A relational function maps its input to a Boolean. That is, it determines if a condition holds for a given input. High school students are commonly familiar with \texttt{$<$}, \texttt{$\leq$}, \texttt{$>$}, and \texttt{$\geq$} when used with specific numbers. They are less comfortable with such functions when variables are involved. The task of determining if a given rocket has consumed a given fuel provides the opportunity to reinforce the importance of relational functions.

\begin{figure}[t]
\begin{center}
  \begin{tikzpicture}
    \draw[loosely dotted] (-2,0) -- node[below] {\tiny W/2} ++(2,0) -- node[below] {\tiny W/2} ++(2,0);
    \draw[loosely dotted] (0,-2) -- node[left] {\tiny H/2} ++(0,2) -- node[left] {\tiny H/2} ++(0,2);
    \filldraw[black] (0,0) circle (2pt) node[anchor=north] {\scriptsize (x,y)};
    \draw[black] (-2,-2) rectangle (2,2);
    %\filldraw[black] (2,2) circle (2pt) node[anchor=west] {\scriptsize (x+W/2,y+H/2)};
    %\draw[dotted,black] (0,0) -- (2,2);
    \filldraw[black] (1.5,0.5) circle (2pt) node[anchor=north] {\scriptsize (x$_{r_1}$,y$_{r_1}$)};
    \filldraw[black] (-1.8,-1.9) circle (2pt) node[anchor=north] {\scriptsize (x$_{r_2}$,y$_{r_2}$)};
    \filldraw[black] (-1.8,-1.9) circle (2pt) node[anchor=north] {\scriptsize (x$_{r_2}$,y$_{r_2}$)};
    \filldraw[black] (3,1.5) circle (2pt) node[anchor=west] {\scriptsize (x$_{r_3}$,y$_{r_3}$)};
    \filldraw[black] (-4,0.6) circle (2pt) node[anchor=south] {\scriptsize (x$_{r_4}$,y$_{r_4}$)};
    \draw (0,0) -- (1.5,0.5);
    \draw (0,0) -- (-1.8,-1.9);
    \draw (0,0) -- (3,1.5);
    \draw (0,0) -- (-4,0.6);
\end{tikzpicture}
\end{center}
  \caption{Illustration of using distance to determine fuel consumption.}\label{dist}
\end{figure}
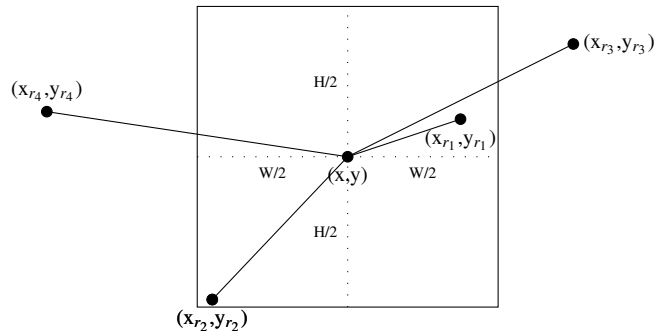

Students are asked how do they know that a rocket has consumed fuel. Most answer that fuel is consumed when the rocket is over the fuel. When pressed on what ``over" means, they state that the image of the rocket is over the image of the fuel. They are unable, however, to state how this can be determined by other than visual inspection (which our computers can't do). At this point, the image displayed in Figure \ref{dist} is piecemeal-built for them. Students are first encouraged to think abstractly of the data definition of a piece of fuel and the image that represents a piece of fuel. The fuel is represented by its center point, \texttt{(x,y)}. The image of a piece of fuel defines a bounding rectangle around the center point. On this image, the center points of a few rockets are drawn (e.g., \texttt{($x_{r_1}$,$y_{r_1}$)}, \texttt{($x_{r_2}$,$y_{r_2}$)}, \texttt{($x_{r_3}$,$y_{r_3}$)}, \texttt{($x_{r_4}$,$y_{r_4}$)}). Student easily observe fuel is consumed when a rocket is inside the fuel's bounding rectangle (e.g, \texttt{($x_{r_1}$,$y_{r_1}$)} and \texttt{($x_{r_2}$,$y_{r_2}$)}). Fuel is not consumed when the rocket is outside the fuel's bounding rectangle (e.g., \texttt{($x_{r_3}$,$y_{r_3}$)} and \texttt{($x_{r_4}$,$y_{r_4}$)}). With this new knowledge, students are once again asked how fuel consumption can be detected. Most, if not all, are still unable to provide any insight on how to solve the problem. After some class discussion, students come to the conclusion that the rocket can be at most half the fuel's image width (i.e., \texttt{W/2}) away on the x-axis and half the fuel's image height (i.e., \texttt{H/2}) away on the y-axis.

At this point, students realize the \emph{relation} that must exist between a rocket and a fuel to conclude that the fuel is consumed by the rocket: the x-distance between the rocket and the fuel must be less than or equal to half the fuel image's width \textbf{and} the y-distance must be less than or equal to half the fuel image's height.

\begin{figure}[t]
  \begin{center}
    \begin{alltt}
      (define HALF-FUEL-IMG-WIDTH (/ (image-width FUEL-IMG) 2))

      (define HALF-FUEL-IMG-HEIGHT (/ (image-height FUEL-IMG) 2))

      ; Sample Expressions
      (define EATEN
         (and (<= (distance-on-x (make-posn 100 340) (make-posn 105 335))
                  HALF-FUEL-IMG-WIDTH)
              (<= (distance-on-y (make-posn 100 340) (make-posn 105 335))
                  HALF-FUEL-IMG-HEIGHT)))

      (define NOTEATEN
         (and (<= (distance-on-x (make-posn 25 10) (make-posn 500 450))
                  HALF-FUEL-IMG-WIDTH)
              (<= (distance-on-y (make-posn 25 10) (make-posn 500 450))
                  HALF-FUEL-IMG-HEIGHT)))

      ; Tests
      (check-expect (eaten? (make-posn 100 340) (make-posn 105 335)) EATEN)
      (check-expect (eaten? (make-posn 25 10) (make-posn 500 450)) NOTEATEN)
      (check-expect (eaten? (make-posn 5 20) (make-posn 4 20)) #true)
      (check-expect (eaten? (make-posn 25 10) (make-posn 320 450)) #false)

      ; rocket fuel --> Boolean
      ; Purpose: To determine if the given rocket has consumed the given fuel
      (define (consumed? a-rocket a-fuel)
         (and (<= (distance-on-x a-rocket a-fuel) HALF-FUEL-IMG-WIDTH)
              (<= (distance-on-y a-rocket a-fuel) HALF-FUEL-IMG-HEIGHT)))
    \end{alltt}
  \end{center}
  \caption{Detecting if a given rocket has consumed a given fuel.} \label{eaten}
\end{figure}

Following a bottom-up approach, the students first develop functions to determine the distance between a rocket and a fuel on the x-axis and the y-axis using the design recipe outlined in Section \ref{Exp}. Focus then turns to using the design recipe to determine if a rocket has consumed a fuel. First, students are asked to pick a couple of sample rockets (say, \texttt{(make-posn 100 340)} and  \texttt{(make-posn 25 10)}) and a couple of sample fuels (say, \texttt{(make-posn 105 335)} and  \texttt{(make-posn 500 450)}) to write sample expressions. With these, students write the sample expressions in Figure \ref{eaten} to complete the first step of the design recipe. In this case, students are made to realize that they need at least two tests: one for each possible outcome (i.e., eaten or not eaten). The reader may observe that the students are using function composition, but they are unaware of this fact. Students are also encouraged to define a constant instead of computing half the fuel's image width and height twice.

After some class discussion, for Step 2 students conclude that there are 2 differences among the sample expressions: the rocket and the fuel. Class discussion is required, because some students think there are 4 differences: the \texttt{x} and \texttt{y} for the rocket and the fuel. That is, students need to be guided to think of a posn as a single value. In this manner, all students are able to realize that the sample expressions are manipulating a rocket and a fuel and not 4 different values. After this realization, for Step 3, students name these differences \texttt{a-rocket} and \texttt{a-fuel}.

Steps 4 and 5 yield the signature and purpose statement displayed in Figure \ref{eaten}. These tend, at this point, to be fairly easy for students to develop. Step 6 is usually more controversial because students take to heart the importance of choosing a function name that helps communicate the purpose of the function. Proposed names include: \texttt{eaten?}, \texttt{consumed?}, \texttt{fuel-consumed?}, \texttt{has-rocket-absorbed-fuel?}. For the sake of brevity, Figure \ref{dist} uses \texttt{consumed?}.

Step 7 is greatly simplified for students, given that they have already chosen sample rockets and fuels and have written sample expressions in Step 1. Students only need to use these, in a consistent manner, to write the first two tests displayed in Figure \ref{eaten}. The second two tests use values not found in the sample expressions. Step 8 has students write the body of the function based on the sample expressions using parameters for the differences.

It is noteworthy that students are usually pleased with how much ``simpler" the body of the function is when compared to the sample expressions. This is a definite win for abstraction. Students also comment that they see why Boolean values and relational functions are important. In addition, students add that they had never thought of, for example, \texttt{$\leq$} or \texttt{>}, as functions. This is an important lesson that students take away from the course.

\subsection{Compound Functions}
Beginning students, overall, are very apprehensive about compound functions. That is, they are uncomfortable with functions that have multiple expressions. This is also true for the high school students in this course. All the students can tell you, for example, what the absolute value of a given number is, but are unable to write it as a function:
\[\
\texttt{absval(x)} =
     \begin{cases}
       \texttt{x} & \texttt{if x} \geq \texttt{0}\\
       \texttt{-x} & otherwise  \\
     \end{cases}
\]
Although most students are able to digest the meaning of the notation above, they feel lost when the function is only slightly more complex. For example, consider the following function:
\[\
\texttt{f(x)} =
     \begin{cases}
       \texttt{x} & \texttt{if x} \leq \texttt{0}\\
       \texttt{x\(\sp{2}\)} &  \texttt{if 0} < \texttt{x} \leq \texttt{5}\\
       \texttt{x+20} & otherwise
     \end{cases}
\]
Students are perplexed by it. Most students are unable to determine, for example, the value of \texttt{f(5)}. This occurs despite the fact that all students can recall studying such functions in their high school Algebra class.

\begin{figure}[t]
 \begin{center}
  \begin{alltt}
  (define MV-UP    (cond [(string=? "up" "up")
                          (move-rocket-up (make-posn 230 315))]
                         [(string=? "up" "down")
                          (move-rocket-down (make-posn 230 315))]
                         [(string=? "up" "left")
                          (move-rocket-left (make-posn 230 315))]
                         [else (move-rocket-right (make-posn 230 315))]))
  (define MV-DOWN  (cond [(string=? "down" "up")
                          (move-rocket-up (make-posn 50 20))]
                         [(string=? "down" "down")
                          (move-rocket-down (make-posn 50 20))]
                         [(string=? "down" "left")
                          (move-rocket-left (make-posn 50 20))]
                         [else (move-rocket-right a-rocket)]))
  (define MV-LEFT  (cond [(string=? "left" "up")
                          (move-rocket-up (make-posn 98 98))]
                         [(string=? "left" "down")
                          (move-rocket-down (make-posn 98 98))]
                         [(string=? "left" "left")
                          (move-rocket-left (make-posn 98 98))]
                         [else (move-rocket-right a-rocket)]))
  (define MV-RIGHT (cond [(string=? "right" "up")
                          (move-rocket-up (make-posn 420 250))]
                         [(string=? "right" "down")
                          (move-rocket-down (make-posn 420 250))]
                         [(string=? "right" "left")
                          (move-rocket-left (make-posn 420 250))]
                         [else (move-rocket-right a-rocket)]))
 \end{alltt}
 \end{center}
\caption{The sample expressions for rocket moving.} \label{mvrocketexps}
\end{figure}

To reinforce understanding of compound functions, students are introduced, for example, to the problem of moving a rocket. To make sure everything is concrete in student minds, a bottom-up approach is used again. That is, first functions to move a given rocket up, down, left, and right are developed in the same manner as outlined in Section \ref{Exp}. Once students have developed these functions, focus is moved onto moving the game's rocket.

Students are taught that compound functions naturally arise when variety in the data being processed is made explicit. For example, for the absolute value function there are two varieties of numbers: negative or nonnegative. Students understand that a rocket's movement depends on the direction value stored in the world. It is explicitly pointed out to students that there is variety in the direction data definition: \texttt{up}, \texttt{down}, \texttt{left}, and \texttt{right}. Therefore, a compound function is needed that first decides how the rocket needs to be moved and then proceeds to move the rocket. At this point, students are introduced to conditional expressions. It is emphasized that conditional expressions are needed to make a decision when processing data whose definition specifies variety.

After several practicing examples writing conditionals to get students familiar with the new syntax, focus returns to writing a function to move a rocket using the design recipe. For Step 1, students must develop at least one example for each direction a rocket can move in. These expressions must decide in what direction to move a rocket to demonstrate how the new rocket is computed. Therefore, students understand that a conditional is needed. Figure \ref{mvrocketexps} displays a sample of such expressions. For Step 2, students observe that the differences are the direction and the rocket. For Step 3, students typically, following course convention, name these differences, respectively, \texttt{a-rocket} and \texttt{a-dir}. It is noteworthy that there may be students that use the same rocket in all their sample expressions and, thus, fail to identify the rocket as a difference. A useful technique to address this problem is to have students repeat Steps 2 and 3 using a fellow student's sample expressions.

\begin{figure}[t]
 \begin{center}
  \begin{alltt}
  (check-expect (move-rocket (make-posn 230 315) "up") MV-UP)
  (check-expect (move-rocket (make-posn 50 20) "down") MV-DOWN)
  (check-expect (move-rocket (make-posn 98 98) "left") MV-LEFT)
  (check-expect (move-rocket (make-posn 420 250) "right") MV-RIGHT)

  (check-expect (move-rocket (make-posn 5 15) "up") (make-posn 5 10))
  (check-expect (move-rocket (make-posn 100 80) "down") (make-posn 100 75))
  (check-expect (move-rocket (make-posn 32 51) "left") (make-posn 27 51))
  (check-expect (move-rocket (make-posn 45 18) "right") (make-posn 50 18))

  ; rocket direction \(\rightarrow\) rocket
  ; Purpose: To move the given rocket in the given direction
  (define (move-rocket a-rocket a-dir)
    (cond [(string=? a-dir "up") (move-rocket-up a-rocket)]
          [(string=? a-dir "down") (move-rocket-down a-rocket)]
          [(string=? a-dir "left") (move-rocket-left a-rocket)]
          [else (move-rocket-right a-rocket)]))
  \end{alltt}
 \end{center}
\caption{The function to move a rocket.} \label{mvrocket}
\end{figure}

Steps 4, 5, and 6 become straightforward for students. They easily realize that the inputs to the function are a rocket and a direction and that the output is a rocket. Students are also able to articulate that the purpose is to move a given rocket in a given direction. The development of the function header also presents no challenges. The results of these steps are displayed in Figure \ref{mvrocket}.

For Step 7, as displayed in Figure \ref{mvrocket}, students write tests using their proposed function. The tests employ the variables defined in Step 1 and employ concrete values not used in Step 1. It is at this point that students realize the value of developing functions on two fronts. First, they realize that using \texttt{move-rocket} is easier than explicitly writing a conditional using concrete values. Second, students begin to appreciate the elegance obtained by defining a function. It is not uncommon for students to comment that it is easier to understand the expressions in the tests that use \texttt{move-rocket} than the conditional in the sample expressions. That is, students appreciate that the development of \texttt{move-rocket} makes their solution to moving a rocket more understandable to others (and to themselves). This is another win for abstraction.

For step 8, students write the body of the function based on their sample expressions and the parameters from Step 3. The result of this step is also displayed in Figure \ref{mvrocket}. The substitution of differences with parameters is straightforward for most students, but instructors would be well-advised to realize that not all students complete the substitution in a timely fashion. This may require coaching some students, but the effort is worthwhile so that no student falls behind. Undergraduate teaching assistants in class have proven invaluable to achieve this. It is also valuable to point out to students that the results of their design suggests that processing different kinds of data requires different kinds of functions. The \texttt{move-rocket} function processes a direction. The \texttt{move-rocket-\{up,down,left-right\}} functions process a rocket.

After successfully running their tests as required by Step 9, it is useful to display for students the \texttt{move-rocket} function using the syntax found in most Mathematics textbooks as\footnote{Due to spacing limitations, \texttt{a-rocket} and \texttt{a-dir} are, respectively, renamed \texttt{r} and \texttt{d}.}:
\[\
\texttt{move-rocket(r, d)} =
     \begin{cases}
       \texttt{move-rocket-up(r)} & \texttt{if d = "up"}\\
       \texttt{move-rocket-down(r)} & \texttt{if d = "down"}\\
       \texttt{move-rocket-left(r)} & \texttt{if d = "left"}\\
       \texttt{move-rocket-right(r)} & otherwise
     \end{cases}
\]
Almost invariably, students begin to understand the syntax used in their high school textbooks and are less intimidated by compound functions. Many students state that they finally understand what functions with a ``big curly bracket" mean.

\subsection{Function Composition}
Many problems are solved using function composition. That is, the output of one function is input to another function. Although junior and senior high school students have studied function composition, an instructor cannot assume that they understand it. This becomes evident when students are unable to explain the following typical notation from high school Algebra textbooks:
\begin{alltt}
          (f \(\circ\) g)(x) = f(g(x))
\end{alltt}
Some students have stated that  \texttt{f}, \texttt{g}, and \texttt{x} are being multiplied. Others have stated that it does not make sense for a function, \texttt{g}, to be input to another function.

In this course, students are focused on the righthand side of the equality. They are asked  what \texttt{g(x)} represents. After some discussion, their answer migrates from a number to a value (i.e., students are guided to remember that the range is not restricted to solely a set of numbers). During class discussion, this value is given a name (e.g., \texttt{g(x) = a}). Therefore, the righthand side of the above equation can be rewritten as \texttt{f(a)}. This now looks familiar to students, and they understand that \texttt{f} is applied to \texttt{a} or, equivalently, \texttt{a} is plugged into \texttt{f}. In this manner, students are led to realize that the output of \texttt{g}, \texttt{a}, is given as input to \texttt{f}.

In order to make this new knowledge relevant, students are told that \texttt{g} computes a specialized value needed by \texttt{f}. If functions are designed using this concept, then their purpose is easier to understand. Instead of computing multiple values within a function, the idea is to identify a specialized value that needs to be computed and use a function to compute said value. This leads to functions that are shorter, easier to design, and easier to understand.

To make this concrete, students are asked to consider writing a function to create an image for a given world. After a short class discussion, students conclude that they need to draw the rocket, the fuel level, the good fuel, and the bad fuel. Adding each of these to a given image can be thought of as computing a specialized value and, therefore, a function is designed for each. Observe that this is, once again, a bottom-up approach that allows students to immediately test a function after it is written.

With the functions for specialized values written, focus turns to designing the function to draw the world. It is not difficult for students to understand that all the elements are drawn over a background image, and they are asked to define variables for the values of sample expressions as required by Step 1 of the design recipe. These are written using sample worlds defined by the students and the specialized drawing functions for the bad fuel, the good fuel, the fuel level, and the rocket. The output of a specialized function is used as input to another specialized function. That is, students are using function composition. These definitions are displayed at the top of Figure \ref{drw}.

\begin{figure}[t]
 \begin{center}
  \begin{alltt}
  (define W1-IMG
          (draw-bfuel (world-bfuel (world-bfuel W1))
                      (draw-gfuel (world-gfuel W1)
                                  (draw-flevel (world-flevel W1)
                                               (draw-rocket W1 BACK-IMG)))))
  (define W2-IMG
          (draw-bfuel (world-bfuel (world-bfuel W2))
                      (draw-gfuel (world-gfuel W2)
                                  (draw-flevel (world-flevel W2)
                                               (draw-rocket W2 BACK-IMG)))))
  (check-expect (draw-world W1) W1-IMG)
  (check-expect (draw-world W2) W2-IMG)
  (check-expect (draw-world (make-world (make-posn 10 10) "right" 8
                                        (make-posn 110 120)
                                        (make-posn 340 170)))
                (draw-bfuel (make-posn 340 170)
                            (draw-gfuel (make-posn 110 120)
                                        (draw-flevel
                                          8
                                          (draw-rocket (make-posn 10 10)
                                          BACK-IMG)))))
  ; world \(\rightarrow\) image
  ; Purpose: To draw the given world in the background image
  (define (draw-world a-world)
    (draw-bfuel (world-bfuel (world-bfuel a-world))
                (draw-gfuel (world-gfuel a-world)
                            (draw-flevel (world-flevel a-world)
                                         (draw-rocket a-world BACK-IMG)))))
  \end{alltt}
 \end{center}
 \caption{The function to draw the world.} \label{drw}
\end{figure}

Students identify the only difference in the sample expressions is the world drawn and name this difference \texttt{a-world} to satisfy Steps 2 and 3 of the design recipe. The signature developed for Step 4 is a fairly straightforward step for students. The development of the purpose statement for Step 5 usually requires some care. Students typically state that the purpose is to draw the given world. This is correct, but incomplete. The purpose statement needs to specify that the world is drawn in the background image. The function header for Step 6 is also fairly straightforward for students. As students develop experience, the tests for Step 7 become easier to develop, and the instructor ought to be able to observe that the steps of the design recipe have been absorbed. The concrete results for each of these steps are displayed in Figure \ref{drw}. Finally, the development of the function's body for Step 8 now only presents a small challenge for students, and most students only discover typing errors when they run the tests for Step 9.
%It is enlightening for students to see the \texttt{draw-world} function that is developed when function composition is not used. Such a function is long, hard to understand, and hard to design. It is especially hard to design, because students are thinking simultaneously about all the subproblems. A sample of such a function is displayed in Figure \ref{drwl} (in Appendix A).

\section{Student Feedback}
\label{Eval}

\pgfplotstableread[row sep=\\,col sep=&]{
cat    & prop \\
1 & 20 \\
2 & 20 \\
3 & 7 \\
4 & 33 \\
5 & 20 \\
}\bund

\pgfplotstableread[row sep=\\,col sep=&]{
cat    & prop \\
1 & 14 \\
2 & 21\\
3 & 29 \\
4 & 29 \\
5 & 7 \\
}\bps

\pgfplotstableread[row sep=\\,col sep=&]{
cat    & prop \\
1 & 0 \\
2 & 13 \\
3 & 47 \\
4 & 13 \\
5 & 27 \\
}\is

\pgfplotstableread[row sep=\\,col sep=&]{
cat    & prop \\
1 & 27 \\
2 & 13 \\
3 & 27 \\
4 & 6 \\
5 & 27 \\
}\intprog

This section presents the empirical data gathered from students. All the students volunteered to take a survey to measure their impressions. The survey included both ordinal and open ended questions. All students answered all questions. On the last day of class, the results of the survey were shared with the students and they were given a chance to comment. In the interest of absolute clarity, it is important to note that due to the small sample size (i.e., 15 students) no general conclusions can be extrapolated for the student population the participants come from.

\begin{figure}[t]
\begin{subfigure}{0.5\textwidth}
\caption{Better Understanding of Functions.} \label{und}
\begin{tikzpicture}
    \begin{axis}[
            ybar,
            symbolic x coords={1,2,3,4,5},
            xtick=data,
        ]
        \addplot[fill=black] table[x=cat,y=prop]{\bund};
    \end{axis}
\end{tikzpicture}
\end{subfigure}%
\begin{subfigure}{0.5\textwidth}
 \caption{Better Problem Solver} \label{ps}
\begin{tikzpicture}
    \begin{axis}[
            ybar,
            symbolic x coords={1,2,3,4,5},
            xtick=data,
        ]
        \addplot[fill=black] table[x=cat,y=prop]{\bps};
    \end{axis}
\end{tikzpicture}
  \end{subfigure}
\caption{Proportions for better understanding of functions and better problem solver.} \label{better}
\end{figure}
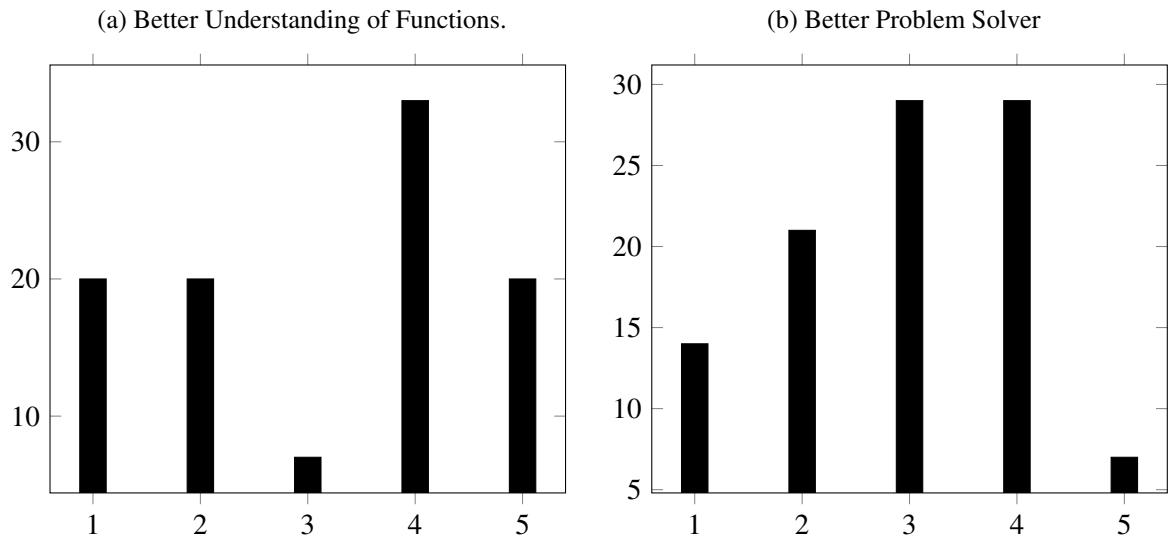

A primary goal of the course is for students to walk away with a better understanding of functions. Students were asked ``\emph{Do you feel you have a better understanding about what functions are now?}" on a scale from 1 (not at all) to 5 (very much so). The distribution of responses is displayed in Figure \ref{und}. The overwhelming majority of students, 60\%, feel strongly that they understand functions better (responses 3--5). This suggests that the course has had a positive impact on students.  A minority of students, 20\%, indicated that they do not have a better understanding of functions (response 1). Invariably, these students expressed that the course should not be required and that they did not wish to be in the course. These feelings are shared by students that responded 2.

Another primary goal of the course is to make students better problem solvers. Students were asked ``\emph{Do you feel you are a better problem solver by knowing how to design and implement
programs?}" on a scale from 1 (not at all) to 5 (very much so). The distribution of responses, displayed in Figure \ref{ps}, suggest that the course has been successful on this front. An overwhelming majority of students, 65\%, strongly feel that they are better problem solvers (responses 3--5). Only 7\% of students felt that they were not better problem solvers (response 1). This is a much lower proportion of students than those that expressed they did not have a better understanding of functions. This suggests that even students that did not wish to be in the course and/or students that did not feel they understood functions better absorbed some problem-solving lessons. It suggests that almost all students, regardless of interest, benefit from exposure to programming.

\begin{figure}[t]
\begin{subfigure}{0.5\textwidth}
 \caption{Interest in Programming.} \label{pint}
\begin{tikzpicture}
    \begin{axis}[
            ybar,
            symbolic x coords={1,2,3,4,5},
            xtick=data,
        ]
        \addplot[fill=black] table[x=cat,y=prop]{\intprog};
    \end{axis}
\end{tikzpicture}
\end{subfigure}
\begin{subfigure}{0.5\textwidth}
\caption{Intellectually Stimulating.} \label{ints}
\begin{tikzpicture}
    \begin{axis}[
            ybar,
            symbolic x coords={1,2,3,4,5},
            xtick=data,
        ]
        \addplot[fill=black] table[x=cat,y=prop]{\is};
    \end{axis}
\end{tikzpicture}
\end{subfigure}%
\caption{Proportions for interest in programming and intellectually stimulating.} \label{interest}
\end{figure}
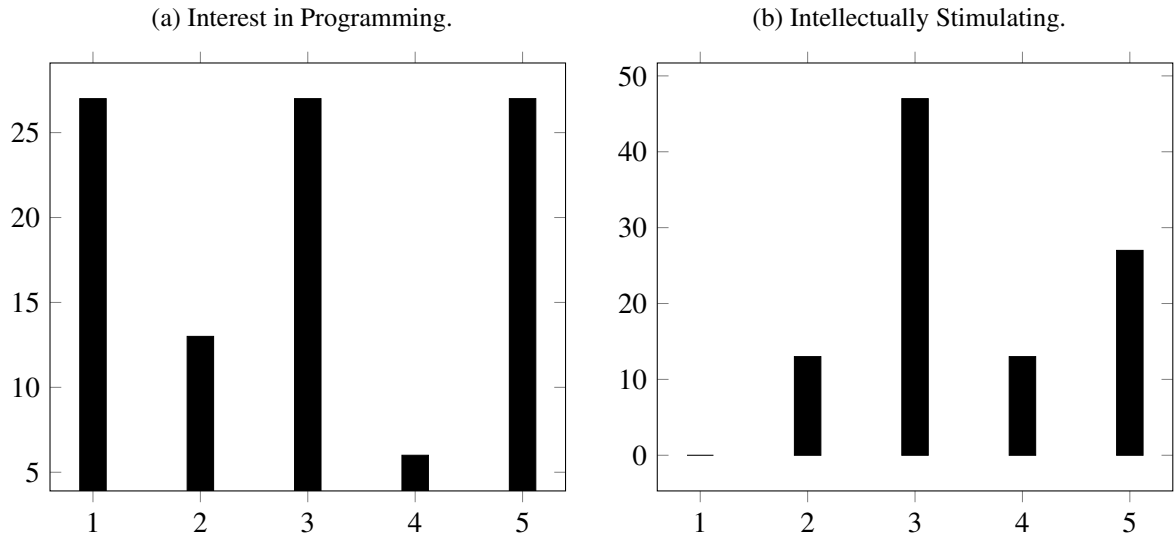

A third primary goal of the course is to spark interest in programming. To measure the effectiveness of the course on this dimension, students were asked ``\emph{What is your level of interest now in programming?}" on a scale from 1 (not at all interested) to 5 (very interested). The distribution of responses in displayed in Figure \ref{pint}. Observe that an overwhelming majority of students, 73\%, express that they are now interested in programming (responses 3--5). This level of interest is a welcomed surprise given that the course started with many obstacles to overcome. There is no doubt that one of the reasons programming became interesting to students is that their education was made relevant. That is, student interest, in part, stems from making programming familiar by using it as a tool to reinforce lessons from high school algebra. This further corroborates the idea that programming ought to be made relevant by using it to teach high school mathematics \cite{CSMATTERS,Schanzer}.

Associating programming with high school algebra sparks interest in programming among students. To further elucidate what exactly is occurring, students were asked ``\emph{How intellectually stimulating is programming?}" on a scale from 1 (not at all stimulating) to 5 (very stimulating). The distribution of responses is displayed in Figure \ref{ints}. An overwhelming majority of students, 87\%, expressed that programming is intellectually stimulating (responses 3--5). This strongly suggests that associating programming and high school algebra is a recipe for success. Students expressed that the course made them think and showed them why algebra is important. They felt the problems being solved were more interesting (unlike ``figuring out when two trains will meet" or ``what are the angles of a triangle").

The open-ended questions asked students what they disliked and liked most about the course and their overall feelings about the course. The most frequent comments about what was disliked are:
\begin{itemize}
  \item \emph{Too much typing}
  \item \emph{Too much work}
\end{itemize}
A full third of the students felt that there was too much typing. This is rather surprising because the developed video game is less than 300 lines (including code, tests, and comments). In open discussion, students in general recognized that this is significantly less than most essays they need to write. Nonetheless, up to one third of the survey answers indicated there was too much typing. A few students felt that overall, the course was too much work. They expressed that they never imagined that writing a video game is so much work. It is important to elucidate why a minority of students felt that 300 lines of code is too much typing and too much work. During open discussion, the majority of students disagreed with the assertion that there was too much typing. In general, the students that felt there was too much typing were the students that had no interest in programming. For example, one student stated that \textit{I want to become a beautician and have no interest in all this typing}. These same students, in general, also felt it was too much work. Students interested in programming or that became interest in programming did not feel the course was too much work or involved too much typing. In fact, they felt that the course was good. Many of these students also expressed wishing that more time were dedicated to this course (instead of other courses they are required to take). They did say that writing sample expressions was a lot of typing at times but appreciated how that led them to the need to define their own functions--precisely the instructor's desired effect. It is also important to note that rarely, if ever, students will comment that a programming course was not too much typing or that the course was not too much work. Simply stated, these are not pressing concerns in the mind of students that have enjoyed the course and have interest in the material. In other words, most students that did not complain about the amount of typing or the amount of work were not even thinking along these lines.

The most frequent comments about what was most liked about the course related to writing the video game, being creative, and having to think. Sample responses from different students include:
\begin{itemize}
  \item \emph{Designing rockets and being creative}
  \item \emph{I liked creating the game because it was fun and it made me think}
  \item \emph{I liked that we were problem solving}
  \item \emph{Everything}
\end{itemize}
These comments further corroborate that students enjoy the personal creative outlet afforded by being able to personalize their games \cite{Yampa,mtm22}. One caveat for instructors is that this enthusiasm must sometimes be reigned in. Many students, given the opportunity, will spend too much time on designing their graphics to the detriment of learning principles of program design.

Finally, different students offered the following comments about their overall feelings about the course:
\begin{itemize}
  \item \emph{I feel it was great. I learned a lot. It is something I might pursue as a major in college}
  \item \emph{It was a very interesting course and I like coding in general}
  \item \emph{I did not like the course because I am not interested in programming at all}
  \item \emph{A lot of typing, but very successful}
\end{itemize}
The above comments confirm the observations reached using the ordinal data. They confirm that the course sparked interest among many. For some, this interest may be strong enough to further pursue learning about programming. There are, of course, students that are simply not interested in programming. It seems, however, that disliking typing did not prevent students from feeling that the course was a success.

\section{Conclusions and Future Work}
\label{Concl}
This article presents a novel approach to introduce high school students to program design and problem solving. The approach is based on high school algebra concepts using a bottom-up approach that allows students to first develop simple functions by abstracting over similar expressions. After students are familiar with abstraction over expressions, they are introduced to developing expressions and functions using high school algebra concepts such as compound functions and function composition. The students are motivated throughout with the development of a functional video game. The empirical data suggests that the course has been successful. Students walk away feeling, for example, that they have a better understanding of functions, that they are better problem solvers, and that they have an interest in programming. This is truly encouraging given that the overwhelming number of students were female, thus, suggesting that it may be an effective approach in closing the Computer Science gender gap.

Future work includes adapting the material for a college-level Computer Science introduction to program design course. The hope is that this approach may be the linchpin needed to make beginning Computer Science students with less than average math skills successful in an introduction to programming course. The approach is also planned to be introduced into a programming course designed for students not studying Computer Science as a major.

Finally, in cooperation with Greg Michaelson \cite{Mich3,Mich2}, we wish to determine if having high school students abstract over expressions that are not formulated using the syntax of a programming language affects how well students can bridge the syntax gap when asked to write code. Our hypothesis is that students that learn how to abstract over expressions first are better equipped to learn how to design and implement programs. This will require a long-term study across several high schools and possibly countries correlating how well learning how to abstract over expressions leads to success in program development.

\section{Acknowledgements}
The author thanks Marva M. Cole-Friday and Abena A. Douglas for administering the \texttt{UBP} program at Seton Hall University and for their unequivocal support during the programming course. The author also thanks Sachin Mahashabde and Jeremy Y. Suero for their invaluable role as teaching assistants.

\bibliographystyle{eptcs}
\bibliography{TFPIE-2020}

\comment{
\section{Appendix A}

\begin{figure}[h]
 \begin{center}
  \begin{alltt}
  (define (draw-world a-world)
    (place-image ; draw bad fuel
      BFUEL-IMG (posn-x (world-bfuel a-world)) posn-y (world-bfuel a-world))
      (place-image ; draw good fuel
         GFUEL-IMG (posn-x (world-gfuel a-world)) (posn-y (world-gfuel a-world))
         (place-image ; draw fuel level
           (rectangle (* (world-flevel a-world) 10) 35 "solid" "purple")
           (- WIDTH 100)
           50
           (cond [(key=? a-dir "up") ; draw rocket
                  (place-image
                    ROCKET-IMG (posn-x (world-rocket a-world))
                    (posn-y (world-rocket a-world))  BACK-IMG)]
                 [(key=? a-dir "left")
                  (place-image
                    (rotate 90 ROCKET-IMG) (posn-x (world-rocket a-world))
                    (posn-y (world-rocket a-world)) BACK-IMG)]
                 [(key=? a-dir "down")
                  (place-image
                    (rotate 180 ROCKET-IMG) (posn-x (world-rocket a-world))
                    (posn-y (world-rocket a-world)) BACK-IMG)]
                 [else (place-image
                         (rotate 270 ROCKET-IMG) (posn-x(world-rocket a-world))
                         (posn-y (world-rocket a-world)) BACK-IMG)])))))
  \end{alltt}
 \end{center}
 \caption{The long version of the function to draw the world.} \label{drwl}
\end{figure}
}

\end{document}